\documentclass[11pt]{article}
\usepackage{epsfig}
\usepackage{graphicx}
\usepackage{epsfig}

\newcommand{\be}{\begin{equation}}

\newcommand{\ee}{\end{equation}}

\newcommand{\bx}{{\bf x}}
\newcommand{\bj}{{\bf j}}

\newcommand{\eq}[1]{(\ref{#1})}
\newcommand{\fig}[1]{Fig.~\ref{#1}}

\newcommand{\cit}[1]{[\ref{#1}]}
\newcommand{\lab}[1]{\protect\label{#1}}

\begin{document}

\title{\bf A variational approach to nonlinear dynamics\\
of nanoscale surface modulations}

\author{V.~B.~Shenoy, A.~Ramasubramaniam and L.~B.~Freund \\
Division of Engineering, Brown University, Providence, RI
02912}
\date{\today}

\maketitle

\begin{abstract}

\narrower{\small In this paper, we propose a variational formulation to study
the singular evolution equations that govern the dynamics of
surface modulations on crystals below the roughening temperature. The basic idea
of the formulation is to expand the surface shape in terms of a complete set of
basis functions and to use a variational principle equivalent to the continuum
evolution equations to obtain coupled nonlinear ordinary differential equations
for the expansion coefficients.  Unlike several earlier approaches that rely on
ad hoc regularization procedures to handle the singularities in the evolution
equations, the only inputs required in the present approach are the orientation
dependent surface energies and the diffusion constants. The method is applied to
study the morphological equilibration of patterned unidirectional and
bidirectional sinusoidal modulations through surface diffusion. In the case of
bidirectional modulations, particular attention is given to the analysis of the
profile decay as a function of ratio of the modulating wavelengths in the
coordinate directions. A key question that we resolve is whether the
one-dimensional decay behavior is recovered as one of the modulating wavelengths
of the two-dimensional profiles diverges, or  whether one-dimensional decay has
qualitatively distinct features that cannot be described as a limiting case of
the two-dimensional behavior. In contrast to some earlier suggestions, our
analytical and numerical studies clearly show that the former situation is true;
we find that the one-dimensional profiles, like the highly elongated
two-dimensional profiles, decay with formation of facets. While our results for
the morphological equilibration of symmetric one-dimensional profiles are in
agreement with the free-boundary formulation of Spohn, the present approach can
also be used to study the evolution of asymmetric profile shapes where the
free-boundary approach is difficult to apply. The variational method is also
used to analyze the decay of unidirectional modulations in the presence of steps
that arise in most experimental studies due to a small misorientation from the
singular surface.  } \medskip

\noindent{\bf Keywords}: Models of nonlinear phenomena, Surface diffusion, Stepped single crystal surfaces

\end{abstract}


\section{Introduction}

In the recent years, nanoscale material structures have been fabricated on solid
surfaces using various patterning techniques which, in some cases, exploit the
phenomenon of stress-driven self-assembly. Potential application of these
structures as functional elements in nano- and micro-electronics has led to
considerable interest in understanding their stability and the processes that
play a role in determining their morphological evolution. Due to their small
size, the time scales involved in the processing and fabrication of these
structures are small. Their small size also implies that the surface energy  is
a significant part of the overall energy of individual structures and, hence, is
important in determining their morphological evolution. For example, transport
of mass via surface diffusion can lead to smaller surface areas, thereby
reducing surface energy for orientation independent surface energy density.
Surface diffusion can also result in transformation of surface orientations with
large surface energies to certain preferred low-energy orientations. Therefore,
a fundamental understanding of the kinetics of surface transport, including the
role of orientation-dependent surface energy, is important in the fabrication of
nanostructures.

In the case of patterned features on the surfaces of amorphous solids or on
crystalline surfaces above the thermodynamic roughening temperature, the surface
energy is independent of the surface orientation. In either situation, surface
diffusion is driven solely by the tendency to reduce the surface area and is
described within the framework established through the classic work of Herring
\cit{Herring} and Mullins \cit{Mullins}. According to their theory, a small
sinusoidal perturbation on a flat surface retains its shape while the amplitude
decays at a rate that is determined by the wavelength of the perturbation. In
the case of a crystal below its thermodynamic roughening temperature, on the
other hand, the situation is more complicated. Because a surface with an
orientation that is close to a high symmetry  orientation is made up of steps,
the dependence of surface energy density on orientation acquires a cusp at such a high symmetry
orientation.

Experiments have shown  that sinusoidal perturbations about these high symmetry
surface orientations do not retain the sinusoidal shape. Instead, the peaks in
the profile become flattened, with the extent of the flat tops increasing as the
amplitude of the profile decays [\ref{Bonzel3}-\ref{Blakely}]. The strong
nonlinearity associated with the cusp in the surface energy causes difficulties
in extending the theory of Mullins to faceted crystals. In this case, the decay
of periodic surface profiles has been theoretically studied using various
continuum models [\ref{Bonzel3}-\ref{Bonzel2},\ref{Murthy1}-\ref{Hager}], by
explicitly accounting for the dynamics of individual steps [\ref{Ozdemir},\ref{Israeli1},\ref{Israeli}]
and by using Monte-Carlo simulations
[\ref{Jiang}-\ref{Erlebacher}]. While useful insights
have been gained, each method has its shortcomings and  the problem of decay
remains open [\ref{Chame}-\ref{Pimpinelli}].  In order to motivate the present
work, which is based on a {\it variational framework} for handling the
nonlinear and singular aspects of surface evolution of crystals below the
roughening temperature, we  briefly summarize here the key results obtained  in
these studies.

The cusp in the surface energy leads to partial differential equations for
surface evolution that involve  Dirac-delta functions or other nonanalytic
functions of the surface slope. In the case of both unidirectional and
bidirectional sinusoidal perturbations, these equations have been solved by
smoothing out the singularities [\ref{Bonzel3}-\ref{Bonzel2},\ref{Murthy1}].
Even though calculations performed using this procedure show flattening of
profile shapes due to facet formation, in agreement experimental observations,
the approach is ad hoc and it is not clear that different smoothing schemes
would lead to identical results. For unidirectional modulations, an alternate
approach was taken by Villain, Zangwill and coworkers
[\ref{Rettori}-\ref{Ozdemir}], who argued that the surface chemical potential
for surface diffusion should not involve the singular term that arises from the
cusp. Retaining only the nonsingular contributions arising from step
interactions, they showed that the sinusoidal profile does not flatten as it
decays but, to the contrary, the top of the profile becomes sharper with time. According to their model, the curvature of the
profile diverges at its extremum points. In distinct contrast to the above
approaches,  Spohn
developed an ingenious method to deal with the singular nature of the evolution
equations for unidirectional modulations by transforming them to a free-boundary
problem [\ref{Spohn},\ref{Hager}]. In his approach, if the location of the facet
is known in advance (which is true for highly symmetric profiles), an ordinary
differential equation for the length of the facet can be coupled to nonsingular
partial differential equations for the evolution of stepped regions to obtain a
coupled problem that is well-posed. Numerical solution of these equations in the
case of unidirectional profiles shows pronounced flattening of the tops
and bottoms of the periodic profiles.

In order to resolve the inconsistent predictions of the various continuum
theories, more atomistic based approaches, particularly Monte Carlo (MC)
simulations [\ref{Jiang}-\ref{Erlebacher}] and models that incorporate the
dynamics of straight steps [\ref{Ozdemir},\ref{Israeli}], have been pursued by
several investigators. While some of the MC simulations show greater tendency to
formation of facets than do others, none of the simulations indicate a
sharpening of the profile as predicted in [\ref{Rettori}] and
[\ref{Ozdemir}]. Since kinetics of surface diffusion within the MC simulations
is slow, most of these simulations have been confined to small sample sizes
which are sensitive to step fluctuations. Furthermore, most of the MC
simulations do not account for the long-range interactions of surface defects
and, therefore, cannot capture the inverse-square elastic or electrostatic
interactions between steps. This problem does not occur in step dynamics
models where step interactions can be included explicitly. However, this approach
also has limitations, since the decay kinetics and profile shapes depend
strongly on the details of the short-range step interactions which are poorly
understood. For example, if an attractive interaction is present between unlike
steps at the top of the sinusoidal profile, the profile tends to be flatter and
decays more rapidly \cit{Israeli}. Also, in the case of bidirectional profiles,
step dynamics models become very complicated and tractable solutions have
been achieved only for cylindrically symmetric surface perturbations that
involve circular steps \cit{Israeli1}; only approximate analytic results for the
case of surface evolution due to evaporation kinetics have been obtained  for
bidirectional sinusoidal perturbations \cit{Lancon}. The more complicated and
experimentally relevant case of surface diffusion limited decay of sinusoidal
profiles has not been treated in a quantitative way using discrete step-dynamics
models.

In this paper, we propose a variational approach to solve the
nonlinear evolution equations for surface evolution and address the problem of
decay of both one- and two-dimensional sinusoidal profiles on crystals below the
roughening temperature. In the past, variational approaches have been used to
study surface evolution involving complex geometries and shapes, but attention
was limited to functional forms of  surface energy that are well-behaved. For
example, a variational model for evolution of a surface with an isotropic surface
energy was introduced by Needleman and Rice \cit{Needleman} to study the problem
of grain boundary transport and cavitation. In recent years, Suo \cit{Suo} has applied the variational formulation to consider surface evolution during motion
of defects such as cracks, voids and inclusions.  The formulation of the
variational approach adopted here is similar,  and
attention is focused on addressing the difficulties posed by nonanalytic
forms of the surface energy.  The basic idea of the formulation is to expand the
surface shape in terms of a complete set of basis functions and to use a
variational principle derived from the continuum evolution equations to obtain
coupled nonlinear ordinary differential equations  for the expansion
coefficients. The advantage of this method is that even though the evolution
equations are singular, the ordinary differential equations that determine the expansion coefficients are
well-behaved and can be integrated by standard numerical techniques. The only
inputs required to solve them  are the orientation-dependent surface energies and
the surface diffusion constants; there are no other ad hoc parameters in the
model.  The conclusions drawn through application of this approach are
briefly summarized here and the details are described in the sections to follow.

In general, the main strength of the variational approach is that it provides a
basis for extracting approximate solutions of a set of equations representing a
physical phenomenon without the need for a priori choices on which features of
those equations to retain or discard.  The approximation made in this approach
is usually the introduction of a lower limit on resolution of features of the
solution, so the approximate solution found is the "best" available within this
restricted range of possible solutions.  The lower limit on resolution can be
reduced indefinitely, in principle, and convergence is expected.  However,
convergence can be proved in only the simplest circumstances, and we are
normally content with establishing essential features of behavior, with the physics of the system
providing guidance on the level of detail needed. As will be demonstrated below,
extensive numerical testing with ever finer resolution is a good indicator of the
level of detail required in any particular problem.

In the case of decay of unidirectional sinusoidal  profiles, when the step
creation energy is finite, our results agree with the predictions of Spohn
[\ref{Spohn},\ref{Hager}]. Here, we find that the decay rate of the amplitude
increases with the  magnitude of step formation energy. On setting the step
creation energy to zero (or if the singular term in the chemical potential is
ignored), we find perfect agreement with the results of Villain and Zangwill
[\ref{Rettori},\ref{Ozdemir}]. This implies that the term arising from the cusp
in the surface energy represents a singular perturbation; the profile decay in
the presence of the this term, however small, is qualitatively different from
the decay when this term is absent. While our model is in agreement with the
free-boundary approach for symmetric profiles, it has several advantages from
both computational and conceptual viewpoints. For example, the free boundary
approach requires the knowledge of the location of facets at the beginning of a
calculation, but this is not known for the case of asymmetric profiles. Within
the variational formulation, this information is not needed a priori and the
approach can therefore be applied to study the evolution of arbitrary profile
shapes. The numerical procedure for solving the free-boundary problem involves
both partial and ordinary differential equations, and  these solutions must be
matched through certain continuity conditions at the edges of the facet. The
present variational formalism involves only solution of coupled ordinary
differential equations and is therefore more straightforward in its
implementation. Perhaps the most significant advantage of the variational
approach is in handling the evolution of two-dimensional profiles, where the
formulation of the free-boundary problem becomes mathematically very involved.
We are not aware of any calculations of two-dimensional decay of sinusoidal
profiles, other than by regularizing the nonanalytic features in the surface
energy. Once again, the present approach does not require ad hoc parameters and
is not fundamentally different from the one-dimensional case.

The ability to handle two-dimensional profiles allows us to address an important
and outstanding issue concerning the limiting behavior of highly elongated two
dimensional profiles, that is, profiles with a wavelength of modulation in one
of the coordinate directions that is much larger than the wavelength in the
other direction. The question we would like to answer is whether the
one-dimensional decay behavior is recovered as one of the modulating wavelengths
of the two-dimensional profiles diverges, or  whether one-dimensional decay  has
qualitatively distinct features that cannot be described as a limiting case of
the two-dimensional behavior. It has been suggested by Rettori and Villian
\cit{Rettori} that the later case is true, who point out the facets form in the
case of two dimensional profiles because of the line tension contribution of
steps in the surface chemical potential. Since there is no contribution of the
step formation energy in their chemical potential for the 1D profiles, they
conclude that there is no facet formation in the 1D case. Within the present
approach, where the line tension contribution of the steps is explicitly
accounted for, we are able to analytically establish that the limiting behavior
of the 2D profiles approach the 1D case. In particular, we show that the
evolution equations of the variational parameters that express the shape of the
2D profiles approach the corresponding equations for the 1D profiles as the
aspect ratio of the 2D profiles (the ratio of wavelengths in the coordinate
directions) approaches infinity. Our analytical results are confirmed by
numerical calculations which show that the 2D profile decay is already close to
the 1D behavior when the aspect ratio is between 100 and 1000.

In most experimental situations, the study of decay of the 1D sinusoidal
perturbations is influenced by the presence of widely spaced steps that run in a
direction perpendicular to the direction of the corrugation. These steps arise
as a result of a very small miscut from the singular orientation and are
difficult to eliminate\footnote{STM images that show the presence of these steps
on patterned Si and Au surfaces are given in \cit{Blakely}.}. In the present
work we will show that, as the miscut angle tends to zero, the decay profiles
approach the shape of the sinusoidal 1D profile on the singular surface. In
contrast to earlier expectations \cit{Lancon}, our studies indicate that the
tendency to form facets persists even as the miscut slope goes to zero.

This paper is organized in the following away. Sec.~2 is devoted to the
development of the variational formulation for both one and two-dimensional
profiles. Here, we will derive evolution equations in the case where mass
transport is controlled by terrace-diffusion. Analytical calculations
establishing the limiting  one-dimensional behavior of highly elongated two
dimensional profiles will also be presented in this section. In Sec.~3, we will
present numerical results for the decay of one-dimensional and two-dimensional
profiles and profiles on surfaces that are miscut with respect to a singular
surface. Particular emphasis is given to the analysis of the decay rate of
two-dimensional profiles as a function of their aspect ratios and the behavior
of sinusoidal profiles on miscut surfaces as the angle of miscut vanishes. In
the case of one-dimensional profiles, we will also consider the decay of asymmetric
profile shapes and present numerical results on the convergence properties of
the variational formulation. Some the predictions of our calculations are
compared with recent experiments \cit{Erlebacker2}. A summary of key results is
given in Sec.~5.

\section{Variational Formulation}

We start by considering the free energy $F$ of the surface $S$, written in terms
of the orientation-dependent surface energy $\gamma$ as
\be
F = \int_S \gamma\left(\nabla_S h\right) dS,
\lab{freeenergy}
\ee
where $h$ denotes the height of the surface measured relative to a high symmetry
reference plane and $\nabla_S$ is the interior gradient operator
on the surface. Using $v_n$ to denote the normal velocity of the
surface relative to the material points instantaneously on that surface, the rate of change
of the total free energy can be expressed in terms of the surface chemical
potential $\mu$ as
\be
\dot{F} = \int_S \mu v_n dS.
\lab{fdot}
\ee
The above equation provides a definition for the surface chemical potential,
which is a surface field representing the driving force for change of shape of
the crystal by diffusive mass transport over its surface. The chemical potential
can be viewed as a continuum representation of the sensitivity of the free
energy of the system to alterations in surface shape. The mass flux on  the
surface, here denoted by the surface vector field ${\bf j}$, is assumed  to be related  to the  surface
chemical potential through a kinetic relationship of the form
\be
{\bf j} = -C \nabla_S \mu,
\lab{kinetic}
\ee
where $C$ is a {\em positive} mobility parameter that, among other things, can
depend on the local surface slope, temperature and the concentration of the
diffusing species.\footnote{Here, for convenience, we assume that surface
diffusion is isotropic, so that the kinetic relationship is written  using
only one mobility parameter. The variational formulation that will be developed
in this section can be generalized to include  diffusion anisotropy in a
straightforward manner.}\ \ Conservation of mass at each point on the surface
requires that the net rate of accumulation of mass at that point on the surface,
determined by the normal speed $v_n$, is the negative of the divergence of the
surface mass flux at that point, or
\be
v_n = - \nabla_S  \cdot {\bf j}.
\lab{massconserve}
\ee
When the surface energy is sufficiently smooth,  eqns
(\ref{freeenergy}-\ref{massconserve}) can be solved by standard numerical
techniques to obtain the evolution of  the
surface shape. Alternatively, the weak form of the equations can be solved  by
employing a finite-element formulation or other variational approaches.
However, if the dependence of a surface energy on surface shape is not analytic, a direct solution of the
evolution equations becomes difficult due to the singularities in the chemical
potential. In such cases, variational approaches provide a powerful tool for determining
the evolution of the surface.

The development of the variational formulation involves two key ingredients.
First, suppose that ${\bf q}$ is any possible mass flux field, sharing
properties  of periodicity and smoothness with ${\bf j}$, but is otherwise
arbitrary. If  we take the inner product of each side of \eq{kinetic} with ${\bf
q}$ and integrate each side over $S$, after the application of a vector
identity, the result can be written in the form
\be
\int_S \nabla_S \cdot (\mu {\bf q}) dS - \int_S \mu \nabla_S \cdot {\bf q} dS + \int_S \frac{1}{C}{\bf j} \cdot {\bf q} dS = 0.
\lab{var2}
\ee
If the fields and the surface shape are spatially periodic and if $S$ includes
exactly one period of the system, then the first term in the above equation
vanishes. This follows from the fact that $\mu$ would have the same value on
opposite sides of a segment of $S$ due to periodicity, and the mass flux out of
$S$ would exactly cancel mass flux out on the opposite side for the same reason.
This result provides the basis for an extremum principle which can be expressed
as a maximum principle or a minimum principle; the two forms are equivalent and
we adopt the latter form here.

Next, define a functional over the range of admissible trial fields ${\bf j}^*$ as
\be
\Phi[{\bf j}^*] = \int_S \left[ -\mu \nabla_S \cdot {\bf j}^* + \frac{1}{2C} {\bf j}^* \cdot {\bf j}^*\right] dS.
\lab{varfun}
\ee
Then \eq{var2} implies that the functional is stationary under arbitrary
variations of the surface flux ${\bf j}^*$ when the flux field is the actual
flux field, that is, when ${\bf j}^* = {\bf j}$. This follows immediately from
\eq{var2} if ${\bf q}$ is interpreted as $\delta {\bf j}^*$ in the usual
variational notation. The arbitrariness of $\delta {\bf j}^*$ and the
fundamental theorem of the calculus of variations require that ${\bf j}$ must
satisfy \eq{kinetic} pointwise on $S$. The fact that $\Phi[{\bf j}^*]$ is not
only stationary but also an absolute minimum under variations about ${\bf j}$
can be seen by considering $\Phi[{\bf j}^*+\delta {\bf j}^*]-\Phi[{\bf j}^*]$ to
second order in $\delta {\bf j}^*$. Since the mobility parameter $C$ is
positive, it follows from \eq{varfun} that
\be
\Phi[{\bf j}^*+\delta {\bf j}^*]-\Phi[{\bf j}^*] = \int_S \frac{1}{2 C} \delta {\bf j}^* \cdot \delta {\bf j}^* dS > 0,
\ee
thus ensuring that ${\bf j}^* = {\bf j}$ is indeed a minimum of $\Phi[{\bf
j}^*]$.

The main advantage of the variational formulation is that instead of solving the
conventional evolution equations, we can expand the mass flux in terms of a
complete set of basis functions and obtain the time dependence of the expansion
coefficients by minimizing the functional $\Phi$. There is no need to deal explicitly with
the singular chemical potential while carrying out this minimization process.
This can be seen by considering the first term in \eq{varfun} which can be seen
to be equal to $\dot{F}$ if we replace the normal velocity $v_n$ in favor of
$-\nabla_S \cdot {\bf j}^*$ in \eq{massconserve}; $\dot{F}$ can be directly
evaluated in terms of ${\bf j}^*$ without using the chemical potential. The
variational functional can then be conveniently expressed as
\be
\Phi[{\bf j}^*] = \dot{F}[{\bf j}^*] + \int_S \frac{1}{2C} {\bf j}^* \cdot {\bf j}^* dS.
\label{varfunctional}
\ee
Below, we provide details of the implementation of the variational formulation
to study the decay of periodic one and two dimensional profiles; the
one-dimensional case is considered first.

\subsection{One-dimensional profiles}

The orientation-dependent surface energy that applies to one-dimensional
modulations can be written as
\be
\gamma(h_x) = \beta_1\left|h_x\right| + \frac{\beta_3}{3}\left|h_x\right|^3,
\lab{1dse}
\ee
where the subscript $x$ denotes the derivative with respect to the spatial
coordinate and the terms proportional to $\beta_1$ and $\beta_3$ represent step
formation and step interaction energies, respectively.  This form of the surface energy
expression presumes that $\vert h_x\vert\ll 1$. If we focus attention on
periodic profiles with wavelength $\lambda$, then free energy in one period and
the position dependent surface chemical potential can be written  as
\be
F = \int_0^{\lambda} \gamma(h_x) dx,
\ee
and
\be
\mu(x) = - \frac{d}{dx}\left(\frac{d \gamma}{dh_x}\right) = -\beta_1\delta\left[h_x\right]h_{xx} - 2\beta_3\left|h_x\right|h_{xx}\,,
\ee
respectively, where $\delta[\,\,]$ denotes the Dirac delta function. The
singular term in the chemical potential can be ignored if the curvature vanishes
at the points where the slope goes to zero. However, since the step interaction
term is cubic in the surface slope, it can be shown that the curvature diverges
like $(x-x_0)^{-1/2}$ around an extremum located at $x = x_0$
[\ref{Rettori},\ref{Ozdemir}]. This implies that the first term plays a crucial
role in determining the decay shape.

If the surface shape is symmetric about the origin, it can be expanded in terms
of a Fourier cosine series as
\be
h(x,t) = \sum_{n=1}^N a_n(t) \cos(nkx), \,\,\,\,\, 0 \le x \le \lambda,
\lab{1dexpansion}
\ee
where $k = 2 \pi/\lambda$ and the number of Fourier modes $N$ is finite but
otherwise unrestricted.\footnote{Numerical studies on the sensitivity of the
results to the number of Fourier coefficients and the convergence properties of
the present formulation will be presented in Sec.~3.}  If attention is focused
on small perturbations that satisfy the criterion $\vert h_x\vert \ll 1$, the normal
velocity of the surface can be written in terms of surface shape as $v_n = h_t$,
where subscript $t$ denotes differentiation with respect to time. We can now
invoke conservation of mass \eq{massconserve} to express the surface
mass flux as
\be
j(x,t) = -\sum_{n = 1}^N\dot{a}_n(t)\frac{\sin(nkx)}{nk}.
\lab{1dmassflux}
\ee
The variational formulation provides first-order coupled ordinary differential
equations for the Fourier coefficients $a_n(t)$; these equations will be
numerically integrated to obtain the evolving surface shape for $t>0$ using
information on the surface shape at $t=0$.

In order to write  the functional $\Phi$ (refer to \eq{varfunctional}) in terms
of the variational parameters, which in the present case are the $\dot{a}_n$s,
we first have to specify the dependence of the surface mobility parameter on the
surface slope. This dependence can be obtained by considering the mechanisms
involved in determining mass transport on stepped surfaces, namely, the
attachment/detachment of atoms from step-edges and the diffusion of adatoms on
the terraces separating the steps. In this paper we will restrict attention to
the case where the mass transport is limited by terrace diffusion;  the mobility
parameter in this situation does not depend the surface slope.\footnote{In case
of attachment-detachment kinetics, a slope dependent mobility parameter
\cit{Nozieres1} can be used to derive the evolution equations. These equations
will be reported elsewhere.}

Using the expression for mass flux in \eq{1dmassflux}, we find the variational
functional for the terrace-diffusion limited case to be
\be
\Phi[\dot{a}_1, \dots, \dot{a}_N] = -\sum_{n = 1}^NH_n[{a}_1, \dots, {a}_N]\dot{a}_n(t) +  \frac{\pi}{2k^3C}
\sum_{n =1}^N \displaystyle{\frac{\dot{a}^2_n }{n^2}},
\ee
where
\be
 H_n[{a}_1, \dots, {a}_N] =  n k \beta_1 \int_0^{\lambda}\mathrm{Sgn}\left[h_x\right]\sin(nkx) dx  +  n k \beta_3\int_0^{\lambda} \mathrm{Sgn}\left[h_x \right] \left|h_x\right|^2\sin(nkx) dx,
\ee
in which Sgn[$p$] $= d|p|/dp$.
Minimizing the functional by setting $\partial \Phi / \partial
\dot{a}_n = 0$ for $n=1,\dots, N$, we obtain the evolution equations
\be
\dot{a}_n(t) = \displaystyle{\frac{n^2 k^3C}{\pi }}H_n[{a}_1 ,\dots, {a}_N]
\lab{evolutioneq}
\ee
for the Fourier coefficients in \eq{1dexpansion}. If the initial shape
is known, approximate solutions to these coupled first-order ordinary
differential equations can be found using standard numerical integration
techniques.

If the problem under consideration is the decay of a single Fourier
mode of amplitude $a_0$, the initial conditions satisfy
\be
a_n(0) = \left \{ \begin{array}{ll}
a_0 & n = 1 \\
0 & {\mathrm {otherwise}}.
\end{array} \right.
\ee
An important point to note is that the evolution equations are well-behaved and
do not directly inherit any singularities that are present in the chemical
potential.  However, the influence of the physical features that give rise
to the singularities are incorporated.

Insights into the nature of the solutions can be gained by expressing
\eq{evolutioneq} in terms of dimensionless variables. Introducing the
rescaled variables
\be
\tilde{h} = kh, \,\,\,\,\,\,\tilde{a}_n = ka_n, \,\,\,\,\,\, \tilde{x} = kx, \,\,\,\,\,\, \tilde{t} =
\displaystyle{\frac{C \beta_3 k^4}{\pi}}t
\ee
we find the Fourier coefficients satisfy
\be
a_n(t) =
 \displaystyle{\frac{1}{k}}\tilde{a}_n
 \left( \tilde{\beta}_1, \tilde{t} \right)
\ee
where $\tilde{\beta}_1 = \beta_1/\beta_3$ and the functional form of the
solution $\tilde{a}_n$ must be determined by solving \eq{evolutioneq}.
We also point out that when the surface shape is not symmetric about the origin,
a complete Fourier expansion that involves both cosine and sine terms should be
adopted; the evolution equations for the expansion coefficients can be derived
by following the steps outlined in this section.

\subsection{Two-dimensional profiles}

The derivation of evolution equations for two-dimensional surface modulations
closely follows the procedure used in the one-dimensional case. In the case of
symmetric profiles, the surface shape can be expressed in terms of the Fourier
series
\be
h(\bx,t) = \sum_{m=1}^N \sum_{n=1}^N A_{mn}(t)\cos(m k_1 x_1)\cos(n k_2 x_2),
\lab{2dshape}
\ee
where $k_{1(2)} = 2\pi/\lambda_{1(2)}$ is the wavenumber of the modulation in
the $x_1(x_2)$-direction. If the amplitude of the modulation is small, we can invoke
conservation of mass, expressed in terms of the surface shape as
\be
\frac{\partial h(\bx,t)}{\partial t} + \nabla \cdot \bj({\bx,t}) = 0,
\ee
to obtain the components of the surface mass flux
\be
(j_1,j_2) = -\sum_m \sum_n\left( \dot{a}_{mn}(t)\frac{ \sin(m k_1 x_1)\cos(n k_2 x_2)}{mk_1}, \dot{b}_{mn}(t)\frac{ \cos(m k_1 x_1)\sin(n k_2 x_2)}{nk_2}\right),
\lab{2dmassflux}
\ee
where
\be
\dot{a}_{mn}(t) + \dot{b}_{mn}(t) = \dot{A}_{mn}(t).
\lab{sum}
\ee

For crystal surfaces that are below the roughening temperature, the  surface
energy can be written as
\be
\gamma(h_{x_1},h_{x_2}) = \beta_1(h_{x_1}^2+h_{x_2}^2)^{1/2} + \frac{\beta_3}{3}(h_{x_1}^2+h_{x_2}^2)^{3/2},
\lab{2dse}
\ee
where $\beta_1$ and $\beta_3$ have the same meaning as in \eq{1dse}. If we
restrict attention to the terrace-diffusion limited surface kinetics, the
mobility parameter in \eq{kinetic} becomes independent of surface slope. In this
case, using \eq{2dmassflux} and \eq{2dse}, the variational functional can be
written as
\be
\Phi[\dot{a}_{11},\dot{a}_{12},\dots,\dot{b}_{11},\dots] = -\sum_{n=1}^N\sum_{m=1}^N(\dot{a}_{mn}+\dot{b}_{mn})H_{mn} + \sum_{n=1}^N\sum_{m=1}^N\frac{\pi^2}{2k_1k_2C}\left[\frac{\dot{a}_{mn}^2}{m^2k_1^2}+\frac{\dot{b}_{mn}^2}{n^2k_2^2}\right],
\ee
where
\be
H_{mn} = \int_0^{\lambda_1}\int_0^{\lambda_2} \left[ m k_1\frac{\partial \gamma}{\partial h_{x_1}}\sin(m k_1 x_1)\cos(n k_2 x_2) + n k_2\frac{\partial \gamma}{\partial h_{x_2}}\cos(m k_1 x_1)\sin(n k_2 x_2) \right]dx_1dx_2.
\ee

Minimizing the functional with respect to $\dot{a}_{mn}$ and
$\dot{b}_{mn}$, we obtain the first-order ordinary differential equations
\begin{equation}
\dot{a}_{mn}(t) = \frac{Cm^2k_1^3k_2}{\pi^2}H_{mn} \,,\quad
\dot{b}_{mn}(t) = \frac{Cn^2k_1k_2^3}{\pi^2}H_{mn}.
\lab{2deveq}
\end{equation}
Adding these two contributions according to \eq{sum}, we find that the evolution
equation for the Fourier amplitude $A_{mn}$ in \eq{2dshape} is
\be
\dot{A}_{mn}(t) = \frac{C(m^2k_1^3k_2 + n^2k_1k_2^3)}{\pi^2}H_{mn}.
\lab{2dfmeveq}
\ee
If the amplitudes of the Fourier modes at $t=0$ are known, \eq{2dfmeveq} can be integrated to obtain the surface shape for
$t >0$.

The dependence of the decay rate on the aspect ratio of the profile
$\alpha \equiv \lambda_1/\lambda_2$ can be understood by using the following dimensionless variables:
\be
\tilde{x}_{1(2)} = k_{1(2)}x_{1(2)}, \,\,\,\, \tilde{h} = {k_1} h, \,\,\,\, \tilde{A}_{mn} = k_1 A_{mn}, \,\,\,\, \tilde{t} = \displaystyle{\frac{C_1 \beta_3 k_1^4}{\pi^2}}t, \,\,\,\, \tilde{H}_{mn} = \displaystyle{\frac{k_2}{\beta_3}} H_{mn}.
\ee
The evolution equations for the scaled Fourier amplitudes in terms of these rescaled variables are
\be
\displaystyle{\frac{d\tilde{A}_{mn}}{d\tilde{t}}} = {(m^2 + \alpha^2n^2)}\tilde{H}_{mn},
\lab{2drseveq}
\ee
where
\be
\tilde{H}_{mn} = \int_{0}^{2 \pi}\int_{0}^{2 \pi}\frac{(\tilde{\beta}_1+\tilde{h}_{\tilde{x}_1}^2 + \alpha^2\tilde{h}_{\tilde{x}_2}^2)}{\sqrt{\tilde{h}_{\tilde{x}_1}^2+\alpha^2\tilde{h}_{\tilde{x}_2}^2}}
\left[{m\tilde{h}_{\tilde{x}_1}}\sin(m \tilde{x}_1)\cos(n \tilde{x}_2)+{n\alpha^2\tilde{h}_{\tilde{x}_2}}\cos(m \tilde{x}_1)\sin(n \tilde{x}_2)\right]d\tilde{x}_{1}d\tilde{x}_{2}.
\lab{2dhmn}
\ee
As noted earlier, the profile decay for the case of $\beta_1 = 0$ is
qualitatively different from the case when this parameter is non-zero but
arbitrarily small. It is also of interest to examine the limiting behavior of
the decay of the two-dimensional profiles as $\alpha \rightarrow 0$, which
corresponds to the situation of one-dimensional behavior determined as a limit
of two-dimensional behavior. To this end, we
proceed to show that the two-dimensional evolution equations  precisely approach
the one-dimensional equations  as $\alpha \rightarrow 0$.

\subsection{Decay of elongated 2D profiles: approach of the 1D limit}
The approach to the one-dimensional limit can be studied by looking at the decay behavior of $h(x_1,0,t)$, given by\footnote{Since the variational approach is strictly exact only when all the Fourier modes are retained, the upper limit in the sum over the Fourier modes will be taken to be infinity. This allows us to use closed form results for certain infinite series; these results are required for establishing the limiting 1D behavior.}
\be
h(x_1,0,t) = \sum_m\left(\sum_{n=1}^{\infty}A_{mn}(t)\right)\cos(mx_1)
\ee
when $\alpha \rightarrow 0$; the goal is to establish that the quantity $A_n^{\infty}(t) \equiv \sum_{m=1}^{\infty}A_{nm}(t)$ obeys the same evolution equations as
the one-dimensional expansion coefficients, $a_n(t)$ in \eq{1dexpansion}. It follows from \eq{2drseveq} and \eq{2dhmn} that when $\alpha \rightarrow 0$, the scaled Fourier amplitudes satisfy
\be
{\frac{d\tilde{A}_{m}^{\infty}}{d\tilde{t}}} = m^2\tilde{H}_{m}^{\infty},
\ee
where
\be
\tilde{H}_{m}^{\infty} =  m\sum_{n=1}^{\infty}\int_{0}^{2 \pi}\int_{0}^{2 \pi}(\tilde{\beta}_1+\tilde{h}_{\tilde{x}_1}^2)
{\mathrm{Sgn}}[{\tilde{h}_{\tilde{x}_1}}]\sin(m \tilde{x}_1)\cos(n \tilde{x}_2)d\tilde{x}_{1}d\tilde{x}_{2}.
\lab{2dhinfty}
\ee
The integral in \eq{2dhinfty} can be explicitly evaluated by noting that
\be
{\mathrm{Sgn}}[{\tilde{h}_{\tilde{x}_1}}] = {\mathrm{Sgn}}[\tilde{h}_{\tilde{x}_1}^{\infty}]{\mathrm{Sgn}}[\cos(\tilde{x}_2)],
\ee
where $\tilde{h}_{\tilde{x}_1}^{\infty} = -\sum_pp\tilde{A}_p^{\infty}\sin(p \tilde{x}_1)$. In what follows we will separately calculate the contribution to $\tilde{H}_{m}^{\infty}$ due to terms that arise from step formation and
interactions.

Using the results
\be
\int_0^{2 \pi}{\mathrm{Sgn}}[\cos(\tilde{x}_2)] \cos(n \tilde{x}_2) d\tilde{x}_2 = \left \{ \begin{array}{ll}
\displaystyle{\frac{4}{n}} & n=1,5,9,\cdots \\
\\
\displaystyle{\frac{-4}{n}} & n=3,7,11,\cdots \\
\\
0 & { n=2,4,6,\cdots}
\end{array} \right.
\ee
and
\be
\sum_{n = 0}^{\infty}\frac{(-1)^n}{(2n+1)} = \frac{\pi}{4},
\ee
the contribution to \eq{2dhinfty} due to the step formation term is shown to be
\be
\tilde{H}_{m}^{\infty}|_f = \pi m\tilde{\beta}_1 \int_0^{2 \pi}{\mathrm{Sgn}}[\tilde{h}_{\tilde{x}_1}^{\infty}] \sin(m \tilde{x}_1)d\tilde{x}_{1}.
\lab{hmif}
\ee
The contributions due to step interactions can be evaluated by employing the expansion
\be
\tilde{h}_{\tilde{x}_1}^2 = \sum_{p,q,r,s}p q \tilde{A}_{pr}\tilde{A}_{qs}\sin(p \tilde{x}_1) \sin(q \tilde{x}_1) \cos(r \tilde{x}_2) \cos(s \tilde{x}_2)
\ee
and the result
\be
\sum_{l=1}^{\infty}\int_0^{2 \pi} {\mathrm{Sgn}}[\cos(\tilde{x}_2)] \cos(l \tilde{x}_2)  \cos(m \tilde{x}_2) \cos(n \tilde{x}_2) d\tilde{x}_2 = \pi,
\lab{3cossum}
\ee
which can be proved for any integers $m$ and $n$ by invoking certain trigonometric identities. Using the fact that the sum in \eq{3cossum} is independent
of both $m$ and $n$, it can be shown that the contribution to \eq{2dhinfty} due to the step interactions is
\be
\tilde{H}_{m}^{\infty}|_i = \pi m \int_0^{2 \pi}(\tilde{h}_{\tilde{x}_1}^{\infty})^2{\mathrm{Sgn}}[\tilde{h}_{\tilde{x}_1}^{\infty}] \sin(m \tilde{x}_1)d\tilde{x}_{1}.
\lab{hmii}
\ee
Combining \eq{hmif} and \eq{hmii} according to \eq{2dhinfty} , we find that
\be
{\frac{d{A}_{n}^{\infty}}{d{t}}} = \frac{C_1 \beta_3 n^3k_1^4}{\pi}\int_0^{\lambda_1}\left(\tilde{\beta}_1+({h}_{{x}_1}^{\infty})^2\right){\mathrm{Sgn}}[{h}_{{x}_1}^{\infty}] \sin(n {x}_1)d{x}_{1},
\ee
which shows that the coefficients ${{A}_{n}^{\infty}}$ satisfy the same
evolution equations as the 1-dimensional expansion coefficients (refer to
\eq{evolutioneq}) if we identify $C_1$, $k_1$ and $\lambda_1$ with $C$, $k$ and
$\lambda$ respectively.  Thus, in the limit of vanishing wavelength aspect
ratio, the two-dimensional case reduces to the independently established one-dimensional
case. Numerical results for profile decay as a function of the
aspect ratio are discussed in the following section.

\subsection{Sinusoidal profiles on vicinal surfaces}

As noted earlier, experiments on 1D sinusoidal modulations are always
 affected by the presence of steps due to small miscuts in directions in a
 direction perpendicular to the direction of the surface modulations. In this
 case of a modulated vicinal surface, the surface height can be expressed as
\be
h(x_1,x_2) = \sum_{n=1}^Na_n(t) \cos(nkx_1) - s_0 x_2,
\lab{vicheight}
\ee
where $\tan^{-1}(s_0)$ is the miscut angle. At time $t=0$, we assume that all the Fourier coefficients except the
longest wavelength mode ($n=1$)have vanishing amplitudes. A schematic of this initial surface shape is given in \fig{vicsin}. We will now present the evolution equations for the Fourier coefficients and discuss their behavior as $s_0 \rightarrow 0$.

Using the procedure used to derive the evolution equations for the 2D profiles, we find the Fourier coefficients satisfy the
evolution equations
\be
\dot{a}_n(t) = \displaystyle{\frac{n^2 k^3C}{\pi }}H_n[{a}_1 ,\dots, {a}_N],
\ee
where
\be
H_n[{a}_1, \dots, {a}_N] =  n k \int_0^{\lambda}\displaystyle{\frac{h_{x_1}}{\sqrt{h_{x_1}^2+s_0^2}}}(\beta_1 +\beta_3 h_{x_1}^2)\sin(nkx_1) dx_1.
\label{hnvicinal}
\ee
It can be seen that these equations have the same form as the evolution equations of the Fourier coefficients
for the 1D profiles given by \eq{evolutioneq},  except for the fact that the
$\mathrm{Sgn}$ function of the surface slope in the expression for $H_n$ is replaced by an analytic function in \eq{hnvicinal}. Furthermore, since
\be
\lim_{s_0 \rightarrow 0}\displaystyle{\frac{h_{x_1}}{\sqrt{h_{x_1}^2+s_0^2}}} = \mathrm{Sgn}[h_{x_1}],
\ee
the evolution equations for the Fourier coefficients in \eq{vicheight} approach the corresponding equations for the
 for the 1D modulations derived earlier. We therefore expect the profile shapes on the miscut surfaces to coincide with the 1D decay profiles of the as the angle of miscut vanishes. The details of the numerical calculations that allow us to
look at this limiting behavior are given in the following section.

\section{Numerical Results}

In this section, we present results on the decay toward morphological equilibrium of
surface modulations obtained by numerically integrating the evolution equations
derived in Sec.~2. We will first consider the decay of one dimensional
sinusoidal perturbations for different values of step-formation and
step-interaction energies. Numerical studies on the sensitivity of the results
to the number of Fourier components that are used to express the surface shape
will then be presented. This is followed by an example of the decay of an
anisotropic profile where the location of the facets are not easily determined
at the beginning of the calculation. In the particular case that we consider,
some of the facets that appear during the early stages of profile evolution,
eventually disappear as other neighboring facets grow at faster rates.  Next, we
consider the decay rates of two-dimensional profiles with  different aspect
ratios and diffusion constants. Finally, we focus on the evolution of sinusoidal
modulations on a vicinal surface that is slightly misoriented with respect to
the flat singular surface.

\subsection{1D sinusoidal profiles}

The evolution equations given by \eq{evolutioneq} were integrated using a
Runge-Kutta integrator with adaptive step-size control discussed in detail in
Ref.~[\ref{Press}]. The time dependence of the amplitude decay, for different
values of rescaled step-formation energies ($\tilde{\beta}_1$),  is shown in
\fig{ampl1D}. The profile shapes at different stages of morphological
equilibration are shown, for certain representative values of $\tilde{\beta}_1$,
in \fig{profiles}. When the step formation energy vanishes, we find that the
profile sharpens at its extrema and the decay rate shows the $1/\tilde{t}$
scaling behavior predicted by Ozdemir and Zangwill \cit{Ozdemir} (refer to inset
(a) in \fig{ampl1D}). On the other hand, for any non-zero value of
$\tilde{\beta}_1$, we find that the profile decays with the formation of facets
which, after an initial transient period of growth, remain nearly constant in size
as seen in \fig{profiles}. The amplitude during this self-similar decay stage,
decreases as a linear function of time, (refer to \fig{ampl1D}), in agreement
with the results of Hager and Spohn \cit{Hager} which were obtained using the
free-boundary approach. With increasing value of $\beta_1$, the lengths of the
facets in the self-similar shape increase in magnitude and the initial transient
periods  decrease in their duration. These results indicate that the case
$\tilde{\beta}_1\neq 0$ is a singular perturbation of the special case
with $\tilde{\beta}_1= 0$; any non-zero value of
$\tilde{\beta}_1$ leads to facet formation during the morphological
equilibration.

The sensitivity of the profile shapes and decay rates on the number of Fourier modes used in the expansion of the surface shape (refer to \eq{1dexpansion}) is considered in \fig{conv1d}, where the decay rates obtained by retaining 15, 20 and 30 modes  are given.  The insets in the figure
show the evolution of the amplitudes of the individual Fourier modes for the three cases considered. We find that the amplitude decay  remains virtually unchanged as the number of Fourier modes increase from 15 to 30. \fig{conv1d} shows that the dominant Fourier components in the surface shape are the long wavelength modes and that the magnitudes of the different Fourier components are also seen to be independent of the total number of modes that are used in the analysis. The insets also show that all the Fourier components decay linearly in time after an initial transient period; in this regime, the surface shape remains self-similar as noted  earlier. We have investigated the time dependence of the decay shapes for several other values of the step-formation energies and initial amplitudes of the sinusoidal profiles and found that accurate results are obtained by using between 15 and 20 Fourier modes. As
is the case with any type of spectral method, if more basis functions are retained, the time steps used in numerical integration have to be decreased, which leads to longer simulation times. The optimum choice for a particular application is determined by a trade-off between accuracy and simulation time.

\subsection{Asymmetric profiles}

Next, we consider the evolution of an asymmetric profile shown in
\fig{arbprof}(a), where the surface shape can be expressed by using both cosine
and sine functions in the Fourier expansion as
\be
h(x,t) = \sum_m a_m(t) \cos(nkx) + \sum_n b_n(t) \sin(nkx).
\lab{cossin}
\ee
The amplitudes of the non-vanishing Fourier components in the initial shape were chosen as follows: $ka_1=0.2, ka_3=0.05, kb_2=0.1$ and $kb_4=0.025$. We find that during the early stages of morphological evolution, facets labeled 1, 2 and 3 appear as shown in
\fig{arbprof}(b). While these facets are situated close to the extrema in the initial profile, their actual locations are not determined very easily. As the shape evolves in time, the facet labeled 3 shrinks in size, while the other two facets grow is size; in \fig{arbprof}(c) this facet has almost disappeared. In the late stages of the evolution, all the Fourier sine components with short wavelengths
vanish, so that the shape closely resembles the decay profile of the symmetric cosine mode with the largest wavelength.

The above example illustrates that the present approach to surface evolution can be used to study arbitrary profile shapes in a straightforward manner. While the free-boundary approach, in principle, can be used in this case, the actual implementation can be quite tedious. First, the locations of the facets are
difficult to guess at the outset and secondly,
application of boundary conditions at the points at which the facets disappear require special consideration. On the other hand, there is no fundamental difference in the analysis of symmetric and asymmetric profiles within the variational approach.

\subsection{2D sinusoidal profiles}

In the case of bidirectional modulations, we start by looking at the decay
behavior of a bidirectional profile whose wavelengths of modulation in the
coordinate directions are equal. The amplitude evolution and profile shapes for
different values of $\tilde{\beta}_1$ are given in \fig{ampl2D} and
\fig{prof11}, respectively. Just as in the 1D case, when $\tilde{\beta}_1 = 0$,
we find that the profile decays without formation of facets.  Furthermore, the
amplitude of the modulation decays with the $1/\tilde{t}$ scaling behavior,
which is also observed in the decay of 1D profiles. It is also evident from
\fig{prof11} that the surface profile close to extrema becomes shaper than the
parabolic shape, which once again resembles the behavior in 1D. For any non-zero
$\tilde{\beta}_1$, the profile decays with the formation of facets, with the
decay rate increasing with $\tilde{\beta}_1$. The decay rate, however, does not
follow the $1/\tilde{t}$ scaling behavior; instead, it shows a linear behavior after
an initial transient as indicated in \fig{prof11}; this trend is also similar to
the decay of a 1D profile. These results indicate that there is no fundamental
difference in the profile decay behavior in one- and two- dimensions; the decay
behavior is governed primarily by the ratio of the step formation and
interaction energies.

In Sec.~2, we presented analytical arguments that establish that the decay of elongated the 2D profiles approaches the 1D limit as the aspect ratio of the profile diverges. We now discuss numerical calculations that allow us to understand the way the profile shapes evolve with increasing aspect ratios. In \fig{conv2D}, the decay rates for profiles with different values of $\alpha$ are given along with the curve for the case $\alpha = 0$, calculated with the 1D evolution equations (refer to \eq{evolutioneq}). We find that as $\alpha$ falls below 1, the decay rates initially decrease  continuously until $\alpha$ reaches about $0.001$. As $\alpha$ decreases further, there is no significant change in the profile decay, so that in the limit of very small $\alpha$ all the decay curves converge to a single curve. It can be seem from
\fig{conv2D} that this curve coincides with the decay profile of the 1D case, in agreement with the analytical results of
Sec.~2. We have also looked at the convergence properties for several other values of  $\tilde{\beta}_1$ and initial amplitudes and have found that the 1D limit is achieved when $\alpha$ lies between 0.01 and 0.001. The evolution of the surface shape during the decay of an elongated profile with $\alpha = 0.1$ is illustrated in \fig{prof110}. It can be seen that well-defined straight features that are aligned along the direction in with the profile is elongated, appear in the decay shape. The cross-section of the profile, particularly at the extrema of the elongated shape, are found to closely resemble the decay profile of a 1D modulation.

\subsection{Sinusoidal profiles on vicinal surfaces}

As a final example of the application of the variational approach, we consider
the decay of sinusoidal profiles on vicinal surfaces that are slightly
misoriented from the flat singular surface. Our goal is to understand the
limiting behavior of profile decay as the miscut slope of the surface approaches
zero. To this end, we consider the decay of a sinusoidal modulation with scaled
amplitude, $ka_1 = 0.1$ in \fig{miscut}. The step formation energy was taken to
be $\tilde{\beta}_1 = 1.0$. The figure shows decay curves for several values of
the misorientation slopes, along with the decay profile for the 1D modulation
calculated using \eq{evolutioneq}. We find that the profile decay progresses at
a faster rate as the misorientation slope decreases from 0.1 to 0.01. As the
slope is decreased further, all the decay curves approach a limiting curve which
coincides with the 1D decay profile. In the particular example that we consider,
the decay behavior is already very close to the 1D case when the misorientation
slope is about $0.001$. We have also numerically investigated the decay behavior
for several other values of step formation energies and initial amplitudes and
found that the decay curves approach the 1D case as the misorientation slope
approaches zero.

Our results differ from the earlier results of Lancon and Villain \cit{Lancon}, who point
out that facets form on surfaces only if the misorientation slope is larger than
some critical value. For the range of parameters considered in \fig{miscut},
they estimate the critical value to be $s_0 = 0.001$. In contrast to their
predictions, our numerical studies clearly show that facets form for arbitrarily
small values of the miscut slope. This observation is consistent with the fact
that the evolution equations for the Fourier coefficients that determine the
shape modulations on  miscut surfaces approach the corresponding equations for
1D coefficients as the miscut slope vanishes.

\subsection{Comparison to experiments on Si(001)}

We now turn our attention to recent experimental work of Erlebacher and coworkers \cit{Erlebacker2}, who investigated the decay of sputter ripples on Si(001) surfaces. They found that amplitude of the ripples, with $\alpha = \lambda_1/\lambda_2 = 0.1$, exhibited a decay behavior consistent with the $1/t$ scaling behavior for 1D modulations predicted by Zangwill and Villain and coworkers [\ref{Rettori},\ref{Ozdemir}]. In this section, using available experimental and theoretical data on step formation and interaction energies for Si(001), we will numerically investigate the decay of these elongated ripples.

The parameters $\beta_1$ and $\beta_3$, for the Si(001) surfaces with alternating SA and SB steps relevant to the experiments, can be estimated from experimentally determined step formation energies reported by Zandvliet \cit{hjwz} and the atomistic calculations of Poon and coworkers \cit{poon}. Using the step formation energies of SA (8 meV/\AA) and SB (16 meV/\AA) steps, we find that $\beta_1 \approx$ 12.5 meV/\AA$^2$. From the atomistic calculations of step interactions on Si(001) \cit{poon}, $\beta_3$ is estimated to be $\approx $1200 meV/\AA$^2$. The profile decay for the elongated ripples with
$\tilde{\beta}_1 = 0.01$ is shown in \fig{smallbeta}. As noted earlier, when $\beta_1 = 0$, both 1D and 2D profiles show a
$1/t$ decay behavior. The figure shows that in distinct contrast to the $\beta_1 = 0$ case, even for a relatively small non-zero value of $\beta_1$, the amplitude decay does not exhibit the $1/t$ scaling behavior. The observed behavior is therefore not consistent with diffusion limited decay of 2D ripples. The results can perhaps be explained by including  anisotropies in surface diffusion and the effects of force-monopoles at the steps due to the 2$\times$1 and 1$\times$2 reconstruction of the terraces between the steps. It has also been suggested that the surface transport at the temperatures
of interest is limited by attachment-detachment kinetics \cit{ikandel}. We leave a detailed consideration of these issues for future study.

\section{Concluding Observations}

In conclusion, we have developed a numerical method to study the singular
equations that govern the dynamics of crystal surfaces below the roughening
temperature. The method was applied to study the morphological equilibration of
both unidirectional and bidirectional modulations. In all these cases, the
surface evolution was determined by solving well-behaved coupled nonlinear
ordinary differential equations for the modal expansion coefficients of the surface
shape. Analysis of highly elongated 2D sinusoidal profiles and sinusoidal
profiles on vicinal surfaces with vanishing miscut angles shows that the
limiting behavior in both of these cases agrees with the 1D decay behavior
accompanied by facet formation. In the following paragraph, we point to
directions of further work that will allow us to make closer connection to
experiments.

In this paper we restricted attention to surface diffusion kinetics limited by
terrace diffusion, while many experimental systems are in a regime where the
mass transport is limited by attachment and detachment at the steps \cit{Blakely}. For the
latter case, Nozieres has derived a kinetic relation between the gradient of the
surface chemical potential and the mass flux \cit{Nozieres1}. In the future we
plan to derive a variational formulation appropriate to this kinetic law to
analyze morphological evolution in the attachment-detachment limited regime.
Also, in some experimental situations, anisotropies associated with surface
diffusion as well as surface energies become can be significant. As noted
earlier, these effects can be included in the present formulation
and will be the subject of forthcoming publications.

\section*{Acknowledgments}

{\small \setlength{\baselineskip}{11pt}{The research support of the National
Science Foundation through grants CMS-0093714 and CMS-0210095 and the Brown University MRSEC
Program is gratefully acknowledged. VBS thanks the
Graduate School at Brown University for providing computational support through
the Salomon Research Award.}}

\section*{References}

\begin{enumerate}

\item \lab{Herring}
C.~Herring,  The use of classical macroscopic concepts in surface energy problems,  {\it Structure and Properties of Solid Surfaces}, edited by R. Gomer and C. S. Smith, University of Chicago Press, (1953) 5.

\item \lab{Mullins}
W.~W.~Mullins, J.~Appl.~Phys. 28 (1957) 333.

\item \lab{Bonzel3}
H.~P.~Bonzel, E.~Preuss and B.~Steffen, Appl. Phys. A 35 (1984) 1.

\item \lab{Bonzel1}
H.~P.~Bonzel, E.~Preuss and B.~Steffen, Surf. Sci. 145 (1984) 20.

\item \lab{Bonzel2}
H.~P.~Bonzel and E.~Preuss, Surf. Sci. 336 (1995) 209.

\item \lab{Umbach}
C.~C.~Umbach, M.~E.~Keeffe and J.~M.~Blakely, J.~Vac.~Sci.~Technol. A 9 (1991) 1014.

\item \lab{Keeffe}
M.~E.~Keeffe, C.~C.~Umbach and J.~M.~Blakely, J.~Phys.Chem.~Solids 45 (1994) 965.

\item \lab{Tanaka1}
S.~Tanaka, C.~C.~Umbach, J.~M.~Blakely, R.~M.~Tromp and M.~Manos, in {\em Structure and Evolution of Surfaces}, Edited
by R.~C.~Cammarata, E.~H.~Chason, T.~L.~Einstein and E.~D.~Williams, MRS Symposium Proceedings, Vol.440 (Materials Research Society, Pittsburg, 1997) p.~25.

\item \lab{Tanaka2}
S.~Tanaka, C.~C.~Umbach, J.~M.~Blakely, R.~M.~Tromp and M.~Manos, J.~Vac.~Sci.~Technol. A 15 (1997) 1345.

\item \lab{Blakely}
J.~Blakely, C.~Umbach and S.~Tanaka, {\em Dynamics of Crystal Surfaces and Interfaces}, Edited by P.~M.~Duxbury
and T.~J.~Pence (Plenum, New York, 1997) p.~23.

\item \lab{Murthy1}
M.~V.~R.~Murthy, Phys.~Rev.~B 62 (2000) 17004.

\item \lab{Rettori}
A.~Rettori and J.~Villain, J.~Phys.~France~49~(1988)~257.

\item \lab{Lancon}
F.~Lancon and J.~Villain, Phys.~Rev.~Lett. 64 (1990) 293.

\item \lab{Ozdemir}
M.~Ozdemir and A.~Zangwill, Phys.~Rev.~B 42 (1990) 5013.

\item \lab{Spohn}
H.~Spohn, J.~Phys.~I ~France 3 (1993) 69.

\item \lab{Hager}
J.~Hager and H.~Spohn, Surf. Sci. 324 (1995) 365.

\item \lab{Israeli1}
N.~Israeli and D.~Kandel, Phys.~Rev.~Lett.~80~(1998) 3300.

\item \lab{Israeli}
N.~Israeli and D.~Kandel, Phys.~Rev.~B~62~(2000) 13707.

\item \lab{Jiang}
Z.~Jiang and C.~Ebner, Phys.~Rev.~B 40 (1989) 316.

\item \lab{Dubson}
M.~A.~Dubson and G.~Jeffers, Phys.~Rev.~B 49 (1994) 8347.

\item \lab{Selke}
W.~Selke and P.~M.~Duxbury, Phys.~Rev.~B 52 (1995) 17468.

\item \lab{Searson}
P.~C.~Searson, R.~Li and K.~Sieradzki, Phys.~Rev.~Lett. 74 (1995) 1395.

\item \lab{Jiang2}
Z.~Jiang and C.~Ebner, Phys.~Rev.~B 53 (1996) 11146.

\item \lab{Murthy2}
M.~V.~R.~Murthy and B.~H.~Cooper, Phys.~Rev.~B 54 (1996) 10377.

\item \lab{Erlebacher}
J.~D.~Erlebacher and M.~J.~Aziz, Surf. Sci. 374 (1997) 427.

\item \lab{Chame}
A.~Chame, F.~Lancon, P.~Politi, G.~Renaud, I.~Vilfan and J.~Villain, Int. Jnl. Mod. Phys. B 11 (1997) 3657.

\item \lab{Bonzel4}
H.~P.~Bonzel, Interface Science, 9 (2001) 21.

\item \lab{Pimpinelli}
A.~Pimpinelli and J.~Villain, {\em Physics of crystal growth}, (Cambridge University Press, Cambridge, 1998) Chap. 8.

\item \lab{Needleman}
A.~Needleman and J.~R.~Rice, Acta~Met.~28~(1980) 1315.

\item \lab{Suo}
Z.~Suo,  { Adv. Appl. Mech.} {33} (1997) 193.

\item \lab{Erlebacker2}
J.~Erlebacher, M.~J.~Aziz, E.~Chason, M.~B.~Sinclair and J.~A.~Floro, Phys.~Rev.~Lett. 84 (2000) 5800.

\item \lab{Nozieres1}
P.~Nozieres, J.~Phys.~France~48~(1987)~1605.

\item \lab{Press}
W.~H.~Press, S.~A.~Teukolsky and W.~T.~Vetterling and B.~P.~Flannery, "Numerical recipes: The art of scientific computing", (Cambridge University Press, Cambridge, 1992) pages 708-715.

\item \lab{hjwz}
H.~J.~W.~Zandvliet,  Rev. Mod. Phys. 72 (2000) 593.

\item \lab{poon}
T.~W.~Poon, S.~Yip, P.~S.~Ho and F.~F.~Abraham, Phys.~Rev.~B 45 (1992) 3521.

\item \lab{ikandel}
N.~Israeli and D.~Kandel, Phys.~Rev.~Lett. 88 (2002) art. no. 169601.
\end{enumerate}
\newpage

\begin{figure}[p]
\vspace{-3cm}
\center{\bf FIGURES}
\begin{center}
\includegraphics[width=5in]{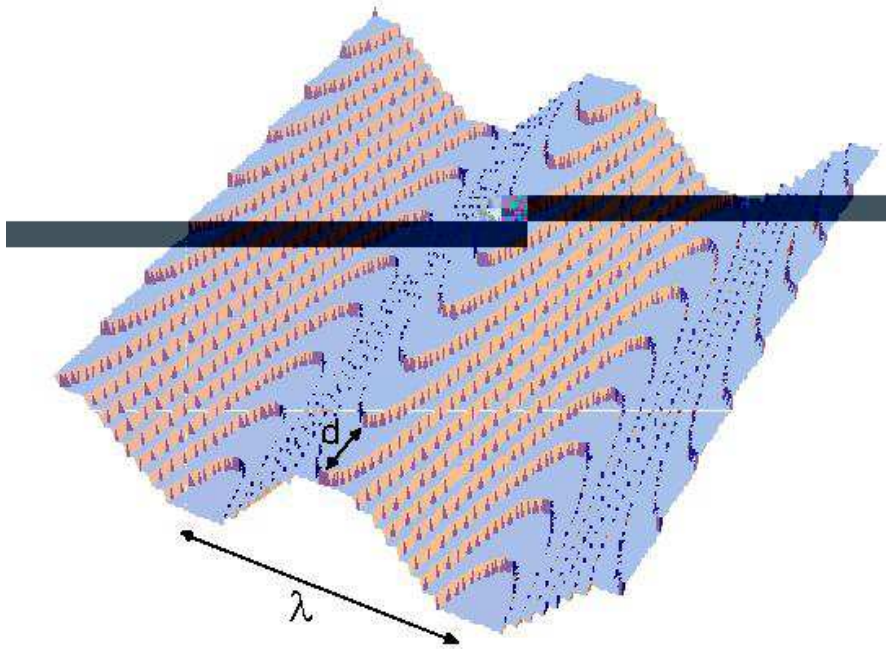}
\end{center}
\vspace{-6cm}
\parbox{1.0cm}{\ }\
\parbox{14cm}{
\caption{ \lab{vicsin} Sketch of a sinusoidal profile on a surface that is slightly misoriented from the flat singular surface. The wavelength of the perturbation is $\lambda$ and the spacing between the steps that arise due to miscut is $d=h_s/s_0$, where $h_s$ is the step height and $s_0$ is the slope of the miscut surface relative to the flat orientation. }}

\end{figure}

\begin{figure}[p]
\vspace{-6cm}
\begin{center}
\includegraphics[width=5in]{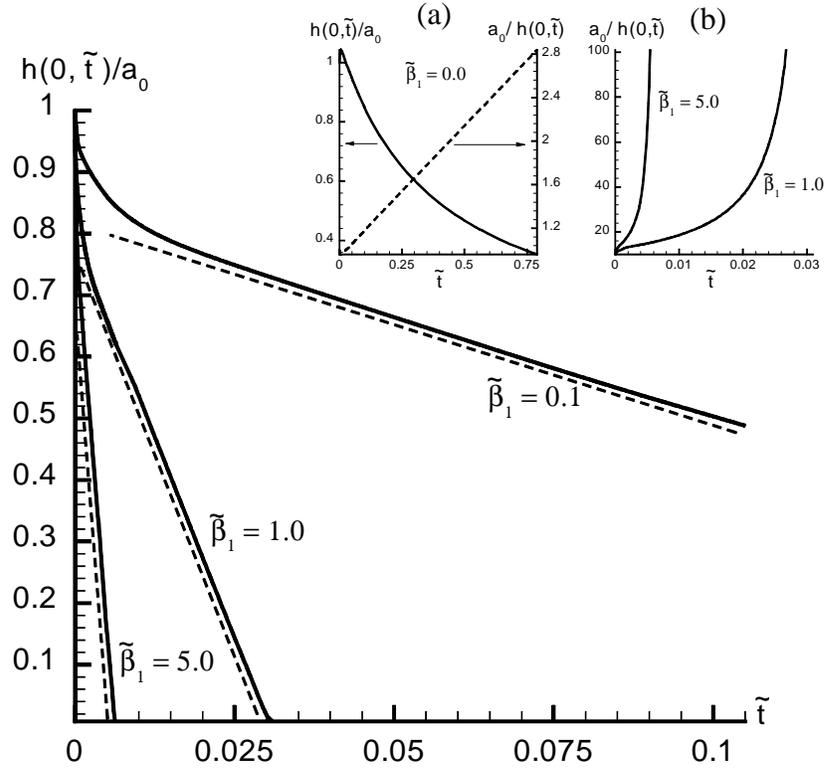}
\end{center}
\vspace{-6cm}
\parbox{1.0cm}{\ }\
\parbox{14cm}{
\caption{\lab{ampl1D} Normalized amplitude $h(0,\tilde{t})/a_0$ as a function
of rescaled time $\tilde{t}$ for the decay of 1D sinusoidal profiles plotted for different values of step-formation energies, $\tilde{\beta}_1$. The initial amplitude of the profile was taken to be
${a}(0) = 0.1/k$, where $k$ is the wavenumber. The inset labeled (a) shows the time dependence of both the normalized amplitude and its inverse for the case of zero step-formation energy ($\tilde{\beta}_1=0$). Here,
amplitude decay is seen to scale as $h(0,\tilde{t})\sim 1/\tilde{t}$, in agreement with the results of \cit{Rettori} and \cit{Ozdemir}. Inset (b) shows that this behavior does not, however, hold for non-zero values of $\tilde{\beta}_1$. The dotted lines next to the decay curves indicate that the amplitudes, after an initial transient period, decay linearly in time.}}
\end{figure}

\begin{figure}[p]
\begin{center}
\includegraphics[width=6.5in]{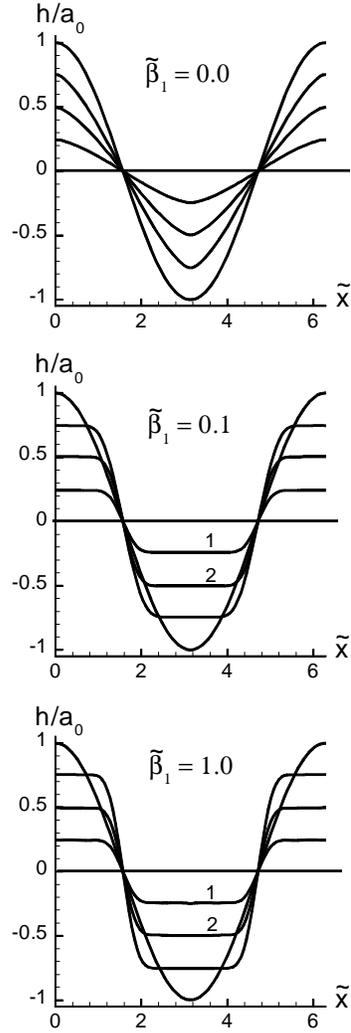}
\end{center}
\vspace{-7cm}
\parbox{1.0cm}{\ }\
\parbox{14cm}{
\caption{\lab{profiles} Morphological equilibration of a symmetric sinusoidal profile for different values of the step-formation energy, $\tilde{\beta}_1$. Note that the propensity to form flat tops increases with increasing values of
$\tilde{\beta}_1$. When $\tilde{\beta}_1 = 0.1$ and $1.0$, the shapes of the curves 1 and 2 marked in the figure are
self-similar; the lengths of the facets remains constant during this decay stage.}}
\end{figure}

\begin{figure}[p]
\begin{center}
\includegraphics[width=5in]{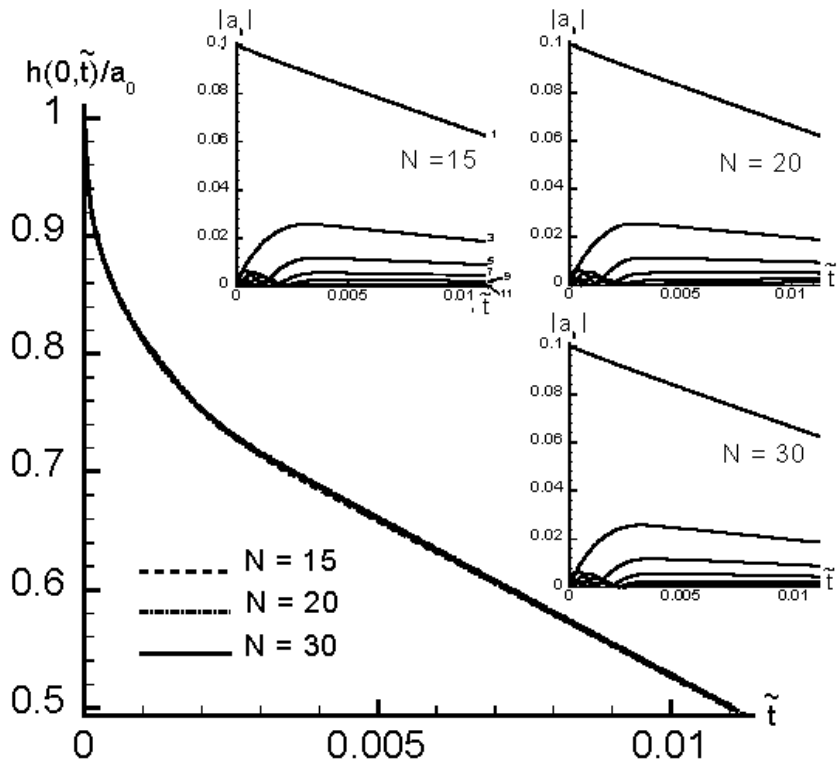}
\end{center}
\vspace{-6cm}
\parbox{1.0cm}{\ }\
\parbox{14cm}{
\caption{\lab{conv1d} Decay of a sinusoidal profile calculated by using 15, 20
and 30 Fourier modes to express the surface shape. The initial amplitude and the
step formation energy(in rescaled units) were taken to be $\tilde{a}_1(0) = 0.1$
and $\tilde{\beta}_1 = 1.0$, respectively. The figure shows that the profile
decay is virtually unchanged as the number of Fourier modes are doubled,
indicating convergence of the method can be achieved by retaining 15-20 modes.
The insets in the figure show the evolution of the amplitudes first 6 non-zero
odd modes for the three cases considered (symmetry arguments can be invoked to
show that even modes do not grow in amplitude). The mode labels are given in the
inset corresponding to N=15. Note that the dominant modes in the profile shapes
are the ones whose wavelengths are large. Also, the amplitudes of the individual
modes remain unchanged as the total number of modes increases from 15 to 30.}}
\end{figure}

\begin{figure}[p]
\begin{center}
\includegraphics[width=5in]{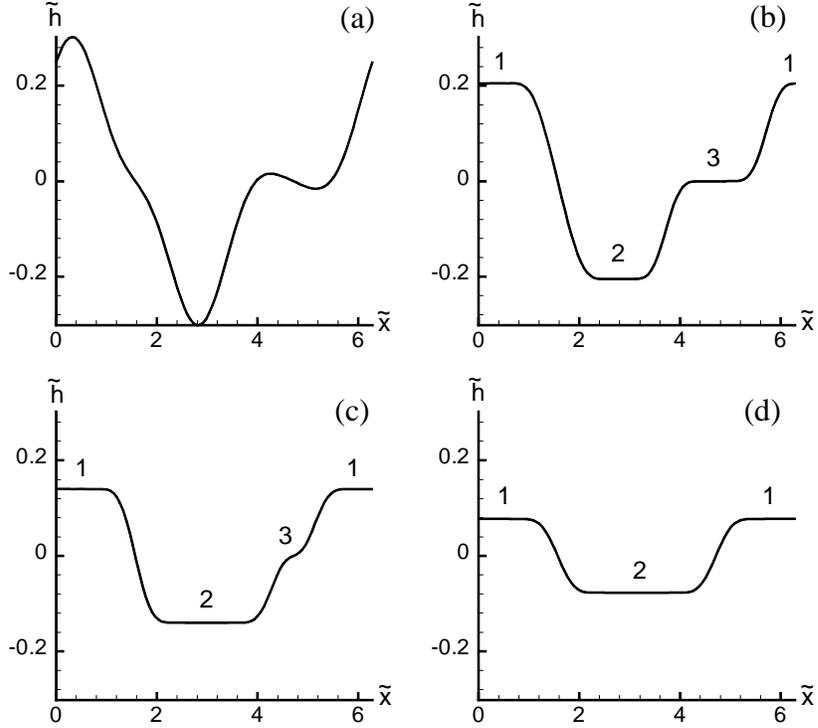}
\end{center}
\vspace{-6cm}
\parbox{1.0cm}{\ }\
\parbox{14cm}{
\caption{\lab{arbprof} The sequence of frames show the time evolution of an
asymmetric profile that contains both cosine and sine modes. The initial profile
shape  is shown in (a); the amplitudes of the non-vanishing modes at $t=0$ are
$ka_1=0.2, ka_3=0.05, kb_2=0.1$ and $kb_4=0.025$ (refer to \eq{cossin}), where
$k$ is the wavenumber of the perturbation. During early stages of decay, three
facets labeled 1, 2 and 3 appear as shown in (b). As the evolution proceeds, the
facets labeled 1 and 2 grow at the expense of 3, which has almost disappeared in
(c). In the late stages of the evolution shown in (d), the shape closely
resembles the decay profile of the symmetric longest wavelength cosine mode. In
this calculation, we have chosen $\tilde{\beta}_1 = 5.0$. The profiles in
(a),(b),(c) and (d) are plotted at $\tilde{t} = 0.0$, $1.9 \times 10^{-6}$, $9.5
\times 10^{-6}$ and $3 \times 10^{-5}$, respectively.}} \end{figure}

\begin{figure}[p]
\begin{center}
\includegraphics[width=5in]{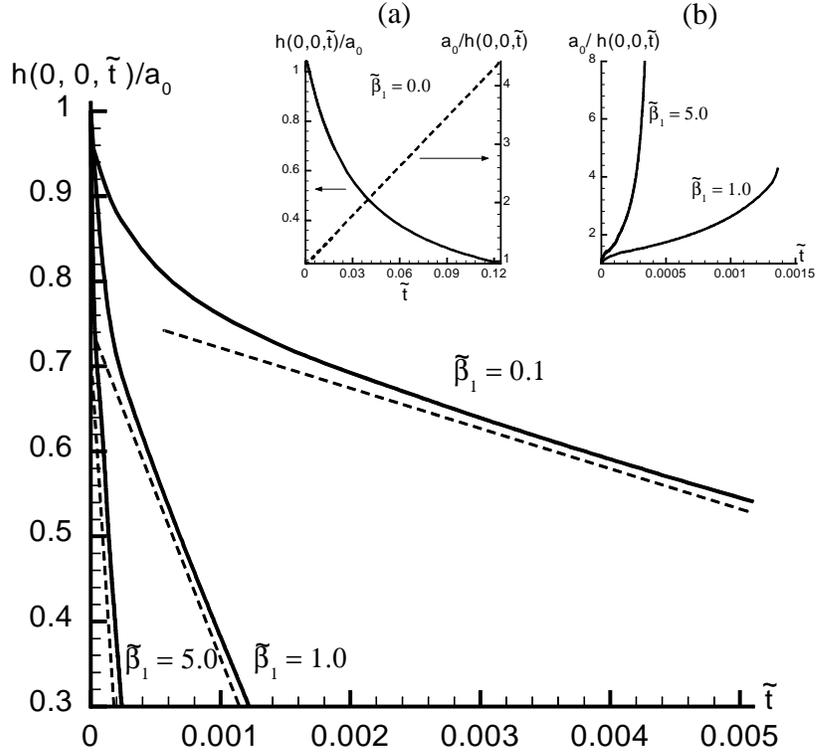}
\end{center}
\vspace{-6cm}
\parbox{1.0cm}{\ }\
\parbox{14cm}{
\caption{\lab{ampl2D} Normalized amplitude $h(0,\tilde{t})/a_0$ as a function
of rescaled time $\tilde{t}$ for the decay of 2D sinusoidal profiles plotted for different values of step-formation energies, $\tilde{\beta}_1$. The wavelengths of the profiles in the coordinate directions are taken to be equal and initial amplitude was assumed to be $a_0 = 0.1/k$, where $k$ is the wavenumber. The inset labeled (a) shows the time dependence of both the normalized amplitude and its inverse for the case of zero step-formation energy ($\tilde{\beta}_1=0$). Here, just as in the 1D case, the amplitude decay is seen to scale as $h(0,\tilde{t})\sim 1/\tilde{t}$. Inset (b) shows that this behavior does not, however, hold for non-zero values of $\tilde{\beta}_1$. The dotted lines next to the decay curves indicate that the amplitudes, after an initial transient period, decay linearly in time.}}
\end{figure}

\begin{figure}[p]
\begin{center}
\includegraphics[width=5in]{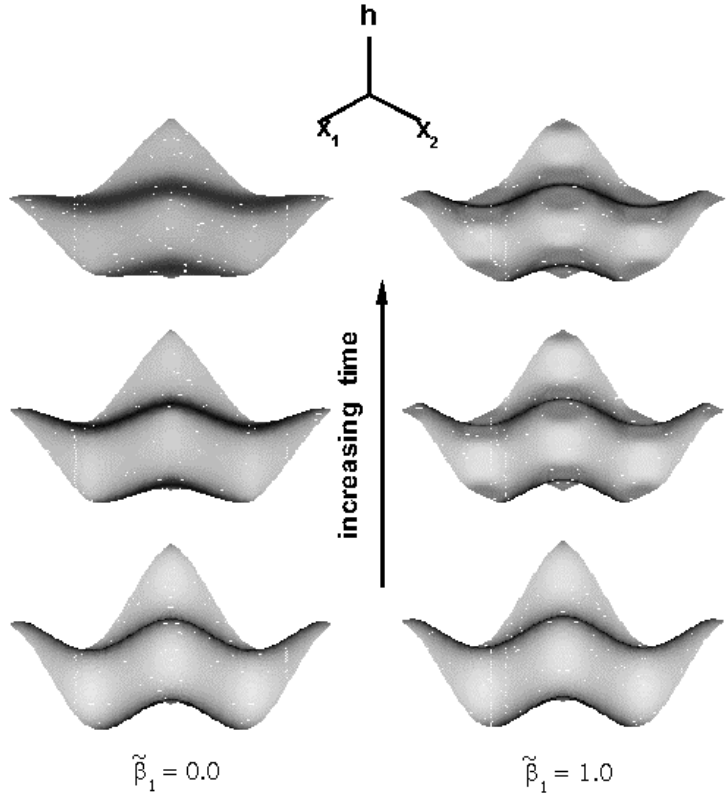}
\end{center}
\vspace{-4cm}
\parbox{1.0cm}{\ }\
\parbox{14cm}{
\caption{\lab{prof11} Time sequence of the decay of bidirectional sinusoidal profiles with $\alpha \equiv \lambda_1/\lambda_2 = 1$, where $\lambda_i$ are the modulating wavelengths in the coordinate directions. The initial amplitude of the modulation is chosen such that $kA_{11} = 0.1$, where $k$ is the wavenumber of the profile. When $\tilde{\beta}_1 =0$, the profile decays without formation of any facets, while facets are formed for any non-zero value of $\tilde{\beta}_1$. For the case $\tilde{\beta}_1 =1.0$, formation of facets at the extrema of the profile is clearly seen in the figure.}}
\end{figure}

\begin{figure}[p]
\begin{center}
\includegraphics[width=5in]{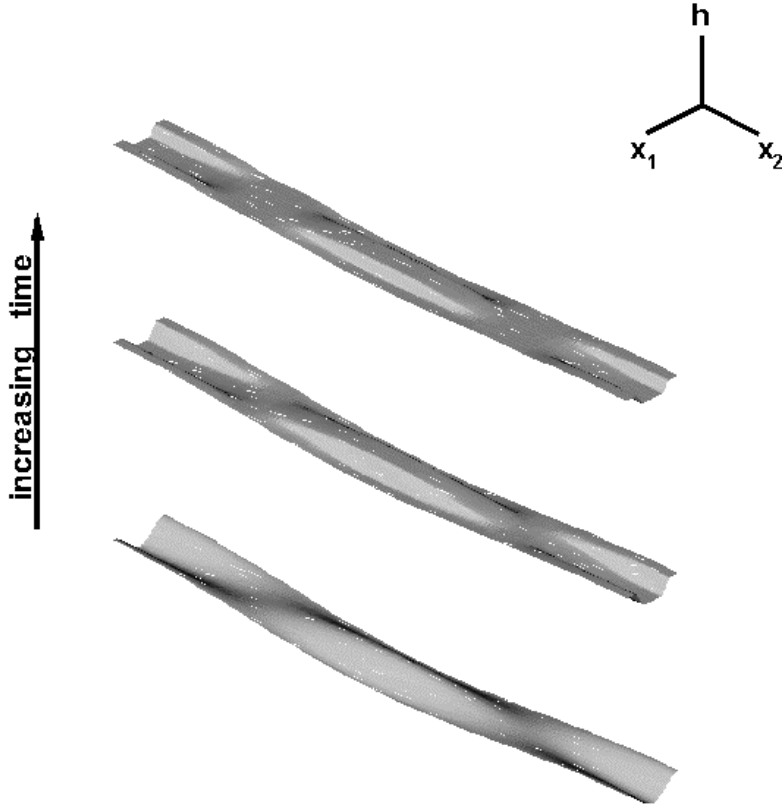}
\end{center}
\vspace{-4cm}
\parbox{1.0cm}{\ }\
\parbox{14cm}{
\caption{\lab{prof110} Time sequence of the decay of a two-dimensional profile with
$\lambda_1/\lambda_2 = 0.1$, where $\lambda_i$ are the modulating wavelengths in the coordinate directions. The scaled step-formation energy is taken to be $\tilde{\beta}_1 = 1.0$ and the initial amplitude is chosen such that $k_1A_{11} = 0.1$, where $k_1$ is the wavenumber along the 1-direction. As the profile evolves, elongated ridges and troughs, aligned along the 2-direction appear in the surface shape. The cross-section of the profile, particularly at the extrema of the elongated shape, is found to closely resemble the decay profile of a 1D modulation.}}
\end{figure}

\begin{figure}[p]
\begin{center}
\includegraphics[width=5in]{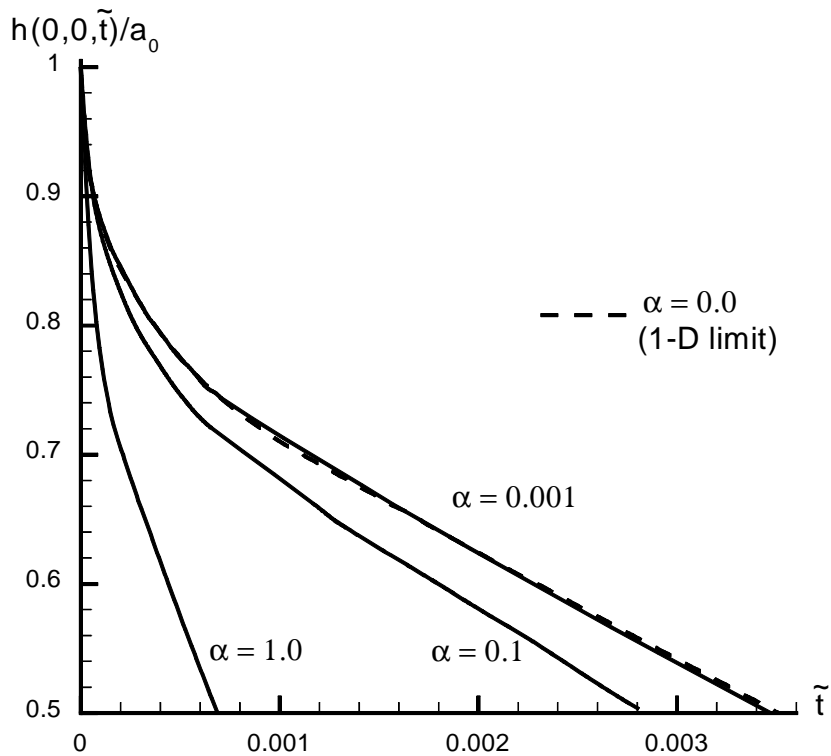}
\end{center}
\vspace{-6cm}
\parbox{1.0cm}{\ }\
\parbox{14cm}{
\caption{\lab{conv2D} Profile decay for bidirectional sinusoidal modulations
plotted for different values of the profile aspect ratios. The aspect ratio is
varied by holding the wavelength  in the $x_1$-direction fixed while increasing
the wavelength of modulation in the $x_2$-direction. The scaled step-formation
energy is taken to be $\tilde{\beta}_1 = 1.0$ and the initial amplitude is
chosen such that $k_1A_{11} = 0.1$, where $k_1$ is the wavenumber along the
1-direction. The figure also shows profile decay for a 1D sinusoidal modulation (dashed-line)
calculated using \eq{evolutioneq}. It can be seen that the decay curve of the
highly elongated profile with $\alpha = 0.001$ very closely follows that for the 1D
case.}} \end{figure}

\begin{figure}[p]
\begin{center}
\includegraphics[width=5in]{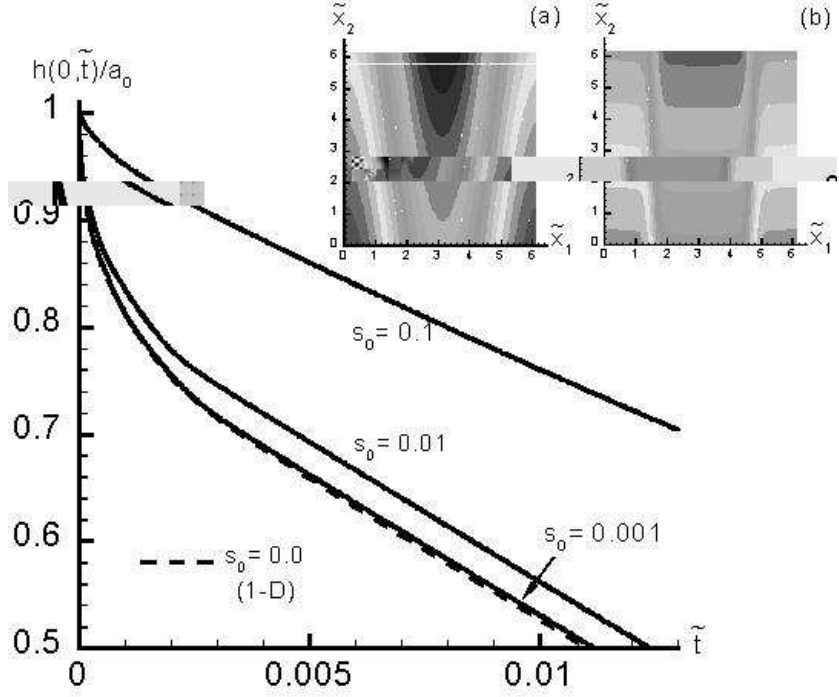}
\end{center}
\vspace{-6cm}
\parbox{1.0cm}{\ }\
\parbox{14cm}{
\caption{\lab{miscut} Profile decay for sinusoidal modulations on surfaces
that are slightly miscut from the flat orientation, plotted for different
values of miscut slopes ($s_0$). The initial amplitude and the step formation
energy (in rescaled units) were taken to be
$\tilde{a}_1(0) = 0.1$ and $\tilde{\beta}_1 = 1.0$, respectively. The figure also shows profile decay for a 1D sinusoidal modulation ($s_0=0$) calculated using \eq{evolutioneq}. Note that the decay curve of the vicinal surface with $s_0 = 0.001$  is seen to almost coincide with the 1D curve. The decay curves for $s_0 < 0.0001$ have not been displayed since they almost exactly coincide with the 1D decay curve. The insets in the figure show the
level curves of surface height (refer to \eq{vicheight}) when $s_0 = 0.001$. The inset labeled (a) shows the level curves of the starting sinusoidal profile, while (b)
shows the level curves after the formation of facets at the extrema of the profile. }}
\end{figure}

\begin{figure}[p]
\begin{center}
\includegraphics[width=6in]{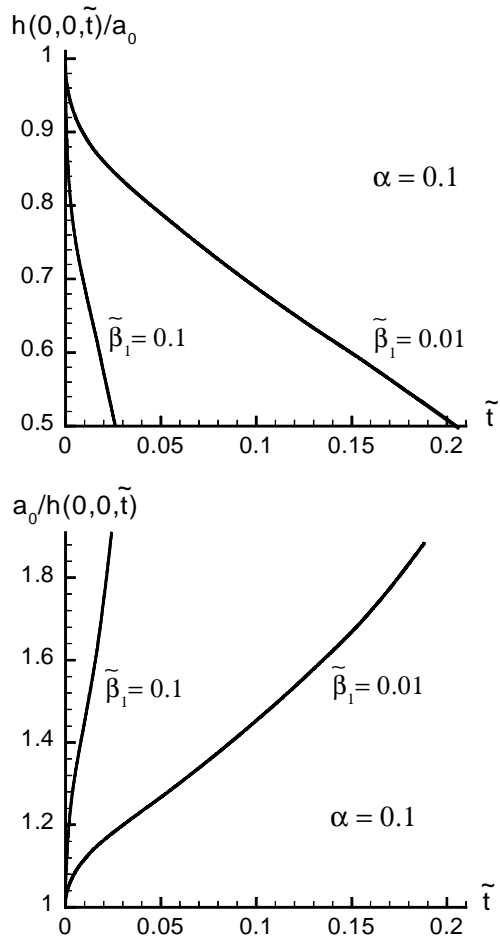}
\end{center}
\vspace{-5cm}
\parbox{1.0cm}{\ }\
\parbox{14cm}{
\caption{\lab{smallbeta} Profile decay of sinusoidal modulations with $\lambda_2/\lambda_1 = 10$, considered in the experimental work of Erlebacher and coworkers \cit{Erlebacker2}. The figure also shows the time dependence of the inverse amplitude for two different values of the scaled step formation energies. In both the cases, the amplitude does not show at the $1/\tilde{t}$ behavior observed in the experiments. Using experimentally determined vales of step formation energies \cit{hjwz}
and step interaction energies obtained form atomistic calculations \cit{poon}, we find that $\tilde{\beta}_1 \approx 0.01$ for SA+SB steps on Si(001). }}
\end{figure}

\end{document}